\documentclass[mathleft]{an}
\usepackage{graphicx}
\usepackage{times}
\begin{document}

% The following seven commands are intended for editorial usage and should be ignored by
% the author(s).
\Pagespan{467}{474}% Document's page range. 
% If second parameter is left empty, the last page is computed automatically.
\Yearpublication{2011}%
\Yearsubmission{2010}%
\Month{5}%   
\Volume{332}%  
\Issue{5}% 
\DOI{10.1002/asna.201011561}% 

\title{Light-element abundance variations in globular clusters\thanks{Highlight talk Astronomische Gesellschaft 2010}}

\author{S.L. Martell\thanks{Corresponding author: \email{martell@ari.uni-heidelberg.de}}}
\titlerunning{Globular cluster abundance variations}
\authorrunning{S.L. Martell}
\institute{Zentrum f\"{u}r Astronomie der Universit\"{a}t Heidelberg, M\"{o}nchhofstrasse 12-14, 69120 Heidelberg, Germany
}

\received{20 December 2010}
\accepted{12 April 2011}
\publonline{26 May 2011}


\abstract{Star-to-star variations in abundances of the light elements
  carbon, nitrogen, oxygen, and sodium have been observed in stars of
  all evolutionary phases in all Galactic globular clusters that have
  been thoroughly studied. The data available for studying this
  phenomenon, and the hypotheses as to its origin, have both
  co-evolved with observing technology; once high-resolution spectra
  were available even for main-sequence stars in globular clusters,
  scenarios involving multiple closely spaced stellar generations
  enriched by feedback from moderate- and high-mass stars began to
  gain traction in the literature. This paper briefly reviews the
  observational history of globular cluster abundance inhomogeneities,
  discusses the presently favored models of their origin, and
  considers several aspects of this problem that require further study.
}

\maketitle

\section{Introduction}
Light-element abundance inhomogeneities in globular clusters have been
a subject of active study for the past thirty years. The phenomenon
was first noticed as anticorrelated variations in the broad CN and CH
absorption features in red giants in a few clusters, and with the
development of larger-aperture telescopes and higher-resolution\linebreak
spectrographs, the data set has expanded in several
directions. Star-to-star abundance variations have been found in stars
from the main sequence to the tip of the red giant branch, the
set of abundances involved has expanded to include carbon, nitrogen,
oxygen, sodium, magnesium and aluminium, and individual studies often survey
tens of clusters rather than one or two. 

Intriguingly, stars with these unusual abundances are\linebreak found
universally in globular clusters, but are apparently only formed in
that environment. The overall picture that has developed is that stars
at all evolutionary phases in all Galactic globular clusters occupy a wide
range in light-\linebreak element abundance that is not observed in any other
Galactic stellar population, or in the fields of Local Group dwarf
galaxies. This abundance range is typically observed in anticorrelated
abundance pairs, C vs N or O vs Na, because roughly half of cluster
stars have abundance patterns like Population II field stars while the
other half are relatively depleted in carbon, oxygen and magnesium and
enhanced in nitrogen, sodium and aluminium, with no variations in any
other elemental abundances.

There are a few globular clusters known to exhibit\linebreak metallicity
variations along with light-element variations, such as $\omega$ Cen
and M54, and they are unusual in other ways as well, with large masses
and likely extragalactic origins (e.g., Carretta et
al. 2010\nocite{CBG10}). However, they are an exception, while
clusters with only light-element variations appear to be the
rule. Comparative studies of light-element abundances in globular
cluster stars and halo field stars (e.g., Langer, Suntzeff \& Kraft
1992\nocite{LSK92}) find that the field star population shows very
little or no light-element abundance variation. The field-star studies
of Pilachowski et al. (1996a)\nocite{PSK96} and Gratton et al. (2000)\nocite{GSCB00} found no stars with cluster-like
abundance variations in the field, and concluded that the cluster
environment must play a vital role in creating or permitting light-element
abundance variations. 

A similar paucity of stars with cluster-like light-element abundances has been
reported in nearby dwarf galaxies.\linebreak McWilliam \& Smecker-Hane (2005)\nocite{MWSH05} and Sbordone et al. (2007)\nocite{SBB07}
both report quite unusual abundance patterns in the Sagittarius dwarf
galaxy, but no stars with cluster-like C-N or O-Na
anticorrelations. Shetrone et al. (2003)\nocite{SVT03} surveyed stars in Sculptor, Fornax,
Carina and Leo I and found none with cluster-like light-element
abudnaces, and Letarte et al. (2010)\nocite{LHT10} confirm that result with more stars in
Fornax. More recently, Maretll \& Grebel (2010)\nocite{MG10} searched a sample of
field giants from the SEGUE survey (Yanny et al. 2009)\nocite{Y09} and\linebreak found that $2.5\%$
of those stars have strong CN and weak CH features relative to the
majority of the field at the same metallicity and luminosity,
suggesting that they may have the full cluster-like light-element
abundance pattern. Those authors claim that the CN-strong field stars
formed in globular clusters and later migrated into the halo as a
result of cluster mass-loss and dissolution processes. 

Studies of Galactic globular cluster abundances are experiencing
something of a revival at present, driven by the discovery that the
(presently) most massive clusters have complex color-magnitude
diagrams (CMDs). There is a variety of unexpected behavior found in
carefully constructed, highly accurate CMDs: multiple main sequences
in M54 (e.g., Siegel et al. 2007\nocite{SDM07}; Carretta et
al. 2010\nocite{CBG10}), $\omega$ Centauri (e.g., Bedin et
al. 2004\nocite{BPA04}; Sollima et al. 2007\nocite{SFB07}; Bellini et
al. 2010\nocite{BBP10}) and NGC 2808 (Piotto et
al. 2007\nocite{PBA07}), multiple subgiant branches in M54, $\omega$
Cen, 47 Tucanae (e.g., Anderson et al. 2009\nocite{APK09}), NGC 1851
(e.g., Cassisi et al. 2008\nocite{CSP08}; Milone et
al. 2008\nocite{MBP08}; Yong \& Grundahl 2008\nocite{YG08}), NGC 6388
(Piotto 2009\nocite{P09}) and M22 (Piotto 2009), and broadened red giant
branches (RGBs) in M4 (Marino et al. 2008\nocite{MVP08}) and NGC 6752
(Milone et al. 2010\nocite{MPK10}). 

There are several tantalizing potential connections between the
photometric multiplicity in clusters and light-\linebreak element abundance
variations. Following the suggestion in Cassisi et al. (2008)\nocite{CSP08} that the two
distinct subgiant branches discovered in NGC 1851 by Milone et al. (2008)\nocite{MBP08}
have very similar ages but total [C+N+O/Fe] abundances that differ by
a factor of two, Yong et al. (2009)\nocite{YGD09} measured abundances of carbon, nitrogen
and oxygen in four bright red giants in NGC 1851. The total [C+N+O/Fe]
abundance in their data set does vary fairly widely, which is
consistent with a model in which NGC 1851 contains two populations in
total light-element abundance. The case of NGC 1851 also provides a
useful illustration of the importance of a careful choice of
photometric systems: while the division of the subgiant branch is
visible in the Milone et al. (2008)\nocite{MBP08} (F336W, F814W) and (F606W, F814W)
color-magnitude diagrams, and in the Johnson-Cousins ($U,I$)
color-\linebreak magnitude diagram presented in Han et al. (2009b)\nocite{HL09}, it is not visible
in the Han et al. (2009b)\nocite{HL09} ($V,I$) color-magnitude diagram or the 
Str\"{o}mgren ($vby$) color-magnitude diagram presented in Yong et al. (2009)\nocite{YGD09}.

In M4, there is also RGB multiplicity that is found or not depending
on the filter set used to construct a color-magnitude diagram:
Marino et al. (2008)\nocite{MVP08} report a split of the red giant branch in the
Johnson-Cousins ($U,B$) color-magnitude diagram, which had not been
observed in previous color-magnitude diagrams based on redder
photometric passbands. However, in M4, the two giant branches appear
to correlate with typical globular cluster variations in [O/Fe] and
[Na/Fe] abundances rather than variations in total [C+N+O/Fe]
abundance. 

It is of course not surprising that abundance variations with
significant spectral effects should have noticeable effects on
photometry, particularly in UV-blue photometric bands where CN and NH
molecular bands can be dominant. Photometric determinations of stellar
parameters and iron abundance were the motivation for defining
medium-band filter systems like Str\"{o}mgren (Str\"{o}mgren 1963)\nocite{S63}, DDO (McClure 1973)\nocite{M73}
and Washington (Canterna 1976)\nocite{C76}. It is, however, unexpected that the
oxygen-sodium anticorrelation\linebreak would affect photometry, particularly in
the very broad\linebreak Johnson/Cousins system, and it is unclear whether
variations in oxygen and sodium abundance, which are observed in all
Galactic globular clusters (e.g., Carretta et al. 2009\nocite{CBG09}),
correlate with variations in $U-B$ color.

The complex phenomenology of photometrically multiple globular clusters,
and the excitement about them, serve to underline the importance of
the rhetorical framework\linebreak used in discussing an observed phenomenon: as
an anomaly, star-to-star abundance variations are a curiosity to be
catalogued, but as a result of light-element self-enrichment and
multiple star-formation events they are powerful markers of the
conditions in early Galactic history. The present challenge is to
understand the connections between the new photometric observations
and the well-studied abundance variations. This requires that we
develop a comprehensive model for the origin of chemical and
photometric complexity in globular clusters, and further that we
ground that model in the larger cosmological environment to enable
studies of the effects of formation time and environment on the ability
of a star cluster to self-enrich and to form a second stellar
generation. 

\section{Development of the observational data set}
Photographic color-magnitude diagrams constructed for\linebreak globular
clusters (e.g., Arp \& Johnson 1955\nocite{AJ55}; Sandage \& Wallerstein
1960\nocite{SW60}) revealed simple stellar populations: unlike the
wide variety found in surveys of the Solar neighborhood, stars in
globular and open clusters all apparently shared a single age and
metallicity. The clusters were\linebreak quickly recognized as ideal
laboratories for testing theories of stellar structure and evolution
(e.g., Sandage 1958\nocite{S58}, Preston 1961\nocite{P61}), and are
used to the present day as anchors for metallicity scales (e.g., Kraft
\& Ivans 2003\nocite{KI03}; Carretta et al. 2009\nocite{CBG09b}).

\subsection{Carbon and nitrogen in bright red giants}
It was therefore suprising when these orderly, predictable stellar
systems turned out to host a number of stars with wide variations in
the strength of molecular features. Unusually weak absorption in the
CH G band (the phenomenon of ``weak-G-band stars'') was noted among
giants in M92 (Zinn 1973\nocite{Z73}; Butler, Carbon \& Kraft
1975\nocite{BCK75}), in NGC 6397 (Mallia 1975\nocite{M75}), in M13 and
M15 (Norris \& Zinn 1977\nocite{NZ77}), in $\omega$ Cen (Dickens \&
Bell 1976\nocite{DB76}), in M5 (Zinn 1977\nocite{Z77}), and in 47 Tuc
(Norris 1978\nocite{N78}). It was quickly shown (Norris \& Zinn 1977;
Zinn 1977) that most of the weak-G-band stars were on the asymptotic
giant branch, implying the existence of some process that dramatically
reshapes the surface abundances of stars between the RGB and the AGB,
described as the ``third dredge-up'' by Iben (1975)\nocite{I75}. 

It was also noted (by, e.g., Norris \& Cottrell 1979\nocite{NC79};
Norris et al. 1981\nocite{N81}; Hesser et al. 1982\nocite{HBH82};
Norris, Freeman \& Da Costa 1984\nocite{N84}) that those RGB stars
with unusually weak CH bands also had relatively strong CN absorption
at 3883\hbox{\AA} and 4215\hbox{\AA}. Since molecular abundance is
controlled by the abundance of the minority species, CH traces carbon
abundance, while CN reflects nitrogen abundance. This general
association of bandstrength and abundance was confirmed by
spectral-synthesis studies such as Bell \& Dickens (1980)\nocite{BD80}, which implies that
the CN-strong, CH-weak stars found\linebreak only in globular clusters have
atmospheres that are depleted in carbon and enriched in nitrogen. 

Since the CNO cycle, operating in equilibrium, tends to convert both
carbon and oxygen into nitrogen, stars with strong CN and weak CH
bands were interpreted as having some amount of CNO-processed material
in their atmospheres. Several theories were put forward to explain
this extra CNO-cycle processing: McClure (1979)\nocite{M79} surveyed the data
available at the time and concluded that internal mixing, specifically
the meridional circulation described by Sweigart \& Mengel (1979)\nocite{SM79}, could be
responsible for ``some or all'' of the surface abundance
variations. The connection between angular momentum and the efficacy
of meridional circulation prompted Suntzeff (1981)\nocite{S81} to propose that
different rotational velocities might explain the different levels of
carbon depletion seen in giants in M3 and M13, an idea further
explored in the Norris (1987)\nocite{N87} study of the relation between CN anomalies
and overall globular cluster ellipticity. Langer (1985)\nocite{L85} suggested that a
uniform mixing efficiency, in combination with star-to-star variations
in CNO-cycle fusion rates, could produce the observed ranges in
surface carbon and nitrogen abundance. Cohen (1978)\nocite{C78} proposed that
star-to-star scatter in [Na/Fe] and [Ca/Fe] in M3 giants required a
non-homogeneous initial gas cloud, a scenario that implies a
primordial origin for [C/Fe] and [N/Fe] variations as well. In
addition, D'Antona et al. (1983)\nocite{D83} proposed that light-element abundance variations
were merely surface pollution, a consequence of mass loss from evolved
stars and the high density of globular clusters. 

\subsection{Other elemental abundances}
Hoping to learn more about the source of apparently CNO-processed
material in the atmospheres of some globular\linebreak cluster giants,
researchers began obtaining higher-\linebreak resolution spectra for cluster
giants. These were difficult observations to make with the 4-m-class
telescopes available at the time, and as a result the data sets were
typically small and limited to stars brighter than $V\simeq
14$. Despite these challenges, it was quickly discovered that stars
depleted in carbon and enhanced in nitrogen were also depleted in
oxygen and magnesium, and enhanced in sodium and aluminium. CN-strong
giants in M5 were found by Sneden et al. (1992)\nocite{S92} to have systematically lower
oxygen abundances and higher sodium abundances than their CN-normal
counterparts. Cottrell \& Da Costa (1981)\nocite{CD81} found positive correlations between CN
bandstrength and both sodium and aluminium abundances in NGC 6752. A
correlation\linebreak between aluminium and sodium abundances, and
anticorrelations between aluminium and both oxygen and magnesium, were
found by Shetrone (1996)\nocite{S96} in M92, M13, M5 and M71, clusters that span the
range of halo globular cluster metallicity. 

Oxygen depletion was consistent with an evolutionary explanation, as
the hydrogen-burning shell of a red giant is hot enough to host the
CNO-cycle reactions $^{14}$N$+2$p$\rightarrow$\linebreak $^{12}$C$+\alpha$. However,
changes in the abundances of heavier elements were unexpected from
fusion reactions occurring within $0.8{\rm M}_{\odot}$ red giants: the NeNa
and MgAl cycles operate similarly to the CNO cycle, with the
nuclei acting as catalysts to convert hydrogen into helium, but both
require significantly higher temperatures. Some authors (e.g.,
Pilachowski et al. 1996b\nocite{P96}) interpreted the extension of the
light-element abundance variations to sodium, magnesium and aluminium
as a sign that the hydrogen-burning shells in red giants must have the
ability to operate the hotter hydrogen-fusion cycles, while others
(e.g., Peterson 1980\nocite{P80}) considered it a sign that the
initial gas cloud from which globular clusters formed must have been
inhomogeneous. 

\subsection{Main-sequence and turnoff stars}
With the construction of 8-meter-class telescopes came access to
fainter stars within globular clusters. Harbeck et al. (2003)\nocite{HSG03} observed around
100 stars at or below the main-sequence turnoff in 47 Tuc, and found
clear bimodality in the distribution of CN bandstrength in those
stars, implying variations in C and N abundance as large as those
already known in red giants in the cluster. Main-sequence and turnoff
stars in M13 were observed by Briley et al. (2004)\nocite{BCS04}, and significant,
anticorrelated ranges in C and N abundance ($\Delta$ [N/Fe] $\simeq
1.0$, $\Delta$ [C/Fe] $\simeq 0.5$) were also found among those stars.

These discoveries had a major impact on theoretical explanations for
light-element abundance variations in globular clusters, and prompted
a serious evaluation of the possibility that they are not simple
stellar populations. While evolutionary explanations could conceivably be
stretched to include modifications of surface aluminium abundance, they
could not accomodate abundance variations in low-mass main-sequence
stars, which are not capable of either high-temperature hydrogen
fusion or mixing between the surface and the core. Additionally, the
fact that the abundance ranges are as large above the ``bump'' in the
RGB luminosity function as below it indicates that the abundance
variations cannot be mere surface pollution, because such a signal
would be greatly diminished at the first dredge-up (Iben 1965)\nocite{I65}, when 
the surface convective zone briefly deepens well into the interior of 
the star. 

\section{Current models for globular cluster formation}
The presently favored explanation for the presence of primordial
light-element abundance variations in globular\linebreak clusters is that the
CN-strong, N- and Na-rich, C- and O-poor stars are a second generation
formed from material processed by intermediate- or high-mass stars in
the first generation. There have been several types of stars proposed
as the source of this feedback material, each with its own strengths
and weaknesses. AGB stars with masses between $4$ and $8 {\rm M}_{\odot}$
(e.g., Parmentier et al. 1999\nocite{PJ99}) are popular because they
are relatively common, they are known to have slow, massive winds,
they are a site of hot hydrogen burning, and they evolve on timescales
of $\sim 10^{8}$ years, quite fast compared to the lifetime of a
globular cluster. In addition to AGB stars, rapidly rotating massive
stars (Decressin et al. 2007b)\nocite{DM07} and massive binary stars undergoing mass transfer (De Mink et al. 2009)\nocite{DM09} have been proposed as alternative sources of feedback
material, and both could deliver more feedback mass in a shorter
amount of time than AGB stars, though the stars themselves are less
common. As is pointed out by Sills \& Glebbeek (2010)\nocite{SG10}, while it may be
observationally determined that one particular feedback source is
dominant, they all certainly contribute to the cluster ISM at some
level. 

In a present-day globular cluster with a mass of $5\times 10^{5}
{\rm M}_{\odot}$ and a $1:1$ ratio of first- to second-generation stars,
assuming a $30\%$ star formation efficiency, the second generation of
stars must have formed from $\simeq 8\times 10^{5} {\rm M}_{\odot}$ of
gas. The first stellar generation, with a mass of $2.5\times 10^{5}
{\rm M}_{\odot}$, clearly cannot have produced enough feedback material to
form the second generation (indeed, typical values for AGB mass loss
are $\le 10\%$). There have been four solutions proposed for this
``mass budget problem'': a top-heavy first-generation mass function
(e.g., Decressin et al. 2007b\nocite{DC07}), a second-generation mass
function that is truncated above\linebreak $0.8 {\rm M}_{\odot}$ (e.g., D'Ercole et
al. 2008\nocite{DVD08}), a first generation that is initially $10-20$ times as
massive as at present (e.g., D'Ercole et al. 2010\nocite{DDV10}), and infall of
pristine gas (e.g., Carretta et al. 2010b\nocite{CBG10b}; Conroy \& Spergel 2010\nocite{CS10}). 

A top-heavy first generation would alleviate the mass budget problem
by placing more $5-15 {\rm M}_{\odot}$ feedback\linebreak sources in the first
generation.  A truncated (bottom-heavy) second generation would
require the first generation to produce less feedback material:
assuming a Kroupa IMF\linebreak (Kroupa et al. 1993)\nocite{KTG93} and a mass range from $0.1$ to
$100 {\rm M}_{\odot}$, half of the mass is in stars with ${\rm M} \le
0.8{\rm M}_{\odot}$. However, it is unclear what physical process would
cause overproduction of massive stars in the first generation or
underproduction in the second, and neither effect is significant
enough to solve the mass budget problem. Additionally, observations of
young star clusters (e.g., Boudreault \& Caballero 2010\nocite{BC10};
Gennaro et al. 2010\nocite{GB10}) do not find either of these effects,
and present-day globular clusters do not contain unusually large
populations of neutron stars or other compact remnants of massive
first-generation stars relative to the number of low-mass
first-generation stars still on the main sequence (e.g., Bogdanov et
al. 2010\nocite{BvH10}; Lorimer 2010\nocite{L10}).
 
 Massive first generations and significant gas infall, in contrast,
 are central to the leading theoretical models of globular cluster
 formation. In the model of D'Ercole et al. (2008)\nocite{DVD08}, the first stellar
 generation has a mass $10$ to $20$ times its present mass, and
 produces all the material needed for the formation of the second
 generation from AGB winds. The authors assume that Type Ia supernovae
 begin occurring 40 Myr after the formation of the first
 generation. The supernovae conclusively end star formation in the
 cluster, meaning that all second-generation stars must have formed
 within that time. The second generation stars form near the cluster
 center, and as a result, when the cluster expands in response to the
 supernovae, the stars that become dissociated from the cluster are
 mostly or exclusively first-generation stars. The authors also
 consider a model with infall of pristine gas from near the globular
 cluster, and find that it prolongs star formation significantly
 beyond the onset of Ia supernovae. 

The globular cluster formation model of Conroy \& \linebreak Spergel (2011)\nocite{CS10} also relies on
AGB stars to provide the\linebreak chemical inhomogeneities between the first
and second stellar generations, but requires significantly more gas
accretion to provide the mass for second-generation stars. The authors
justify this by speculating about the cosmological environment of
proto-globular clusters, namely self-gravitating,\linebreak gas-dominated
proto-galactic systems with gas masses between $10^{8} {\rm M}_{\odot}$ and
$10^{10} {\rm M}_{\odot}$. They find that globular clusters orbiting in such
systems can accrete significant amounts of gas over $10^{8}$ years,
through a combination of Bondi accretion and ``sweeping up'' of
material in the cluster's path, with little vulnerability to
ram-pressure stripping for clusters above $10^{4}{\rm M}_{\odot}$. In this
model, Lyman-Werner flux\linebreak ($912 \hbox{\AA} \le \lambda \le 1100
\hbox{\AA}$) from massive first-generation stars prevents the
gathering gas from forming stars by dissociating H$_{2}$ molecules,
creating a gap of roughly $10^{8}$ years between the two generations
and allowing time for the mass in accreted gas to become large enough
to create a second generation as massive as the first. 

Any globular cluster formation model that requires significant gas
accretion implicitly assumes that the accreted gas must have a
metallicity very similar to that of the first stellar generation,
since systematic [Fe/H] differences are not observed between first-
and second-generation cluster stars. There are a few models that are
explicit about the importance of this coincidence, such as
Carretta et al. (2010a)\nocite{CBG10} and Smith (2010)\nocite{S10}. Both of these models incorporate
supernova material into the second generation but require that it mix
well with accreted lower-metallicity gas before the formation of the
second generation in order to make the metallicity of the second
stellar generation be the same as the first. While these conditions
are certainly conceivable for certain proto-globular clusters, they
are unlikely to hold for all globular clusters in the Milky Way. 

\section{Recent observational progress}
There have recently been several large-scale light-element abundance studies, which have the distinct advantage over smaller-scale studies that the abundance behavior of multiple clusters can be directly compared within homogeneous observations, data reduction and analysis. The low-\linebreak resolution study of Kayser et al. (2008)\nocite{KHG08} measured CN and CH bandstrengths in stars from the main sequence to the red giant branch in 8 Southern globular clusters. Those authors confirmed the presence of RGB CN bandstrength variations and CN-CH anticorrelations, as found in previous single- and few-cluster studies. They also demonstrated that CN bandstrength variations can be found on the main sequence in the clusters NGC 288, NGC 362, M22 and M55, clusters that had not previously been observed to contain main-sequence abundance variations. To explore the question of the source of light-element abundance variations, they also evaluate possible correlations between the ratio of CN-strong to CN-weak stars and several cluster parameters, and find mild positive correlations to cluster luminosity and tidal radius. These trends are interpreted as signs that globular clusters with larger masses or outer-halo orbits would be more efficient at producing second-generation stars.

The comprehensive study of Carretta et al. (2009b)\nocite{CBG09} reported homogeneous oxygen
and sodium abundances for 1958 stars of all evolutionary phases in 19
Southern globular clusters. This study made a significant statement
about the universality of light-element abundance variations in
globular clusters, and also explicitly adopted the language of
self-enrichment and multiple stellar generations. By identifying
groups of stars as ``primordial'', ``intermediate'' or ``extreme''
depending on oxygen and sodium abundances, this study made the claim
that the degree of abundance variation can differ between clusters,
and may be a function of environment and feedback
source. Main-sequence stars in many of the same clusters were observed
by Pancino et al. (2010)\nocite{PR10}, who measured CN and CH bandstrength variations from
low-resolution spectra. These authors found that the fraction of
CN-strong (second-generation) stars was $\sim 30\%$, distinctly lower
than the $70\%$ reported in Carretta et al. (2009b)\nocite{CBG09}. This discrepancy is
curious, and Pancino et al. (2010)\nocite{PR10} suggest that it may indicate that C-N
abundance variations are contributed, at least in part, by a different
feedback source from the O-Na abundance variations studied by
Carretta et al. (2009b)\nocite{CBG09}. 

From recent large-scale studies it appears that light-\linebreak element
abundance variations are universal in Galactic globular clusters, but
the question of the dominant first-\linebreak generation feedback source remains
unsolved. Future studies will need to measure the C-N and O-Na variations
simultaneously in order to address the mismatch in frequency of
second-generation stars found in
low- versus high-\linebreak resolution spectroscopic studies, and will need to measure
other specific elemental abundances to evaluate various aspects of
cluster formation scenarios. For example, abundances of s-process
elements like Ba would be useful for placing limits on AGB
contributions (e.g., Smith 2008\nocite{S08};\linebreak Yong et
al. 2008\nocite{YK08}), and the abundance of Li carries a great deal
of information about the importance of infalling pristine gas (e.g.,
D'Orazi \& Marino 2010\nocite{DM10}; Shen et al. 2010\nocite{SB10}).

The APOGEE survey (Allende Prieto et al. 2008)\nocite{AP08}, one of four
components of the SDSS-III project, will obtain\linebreak high-resolution
near-infrared spectra for 100~000 stars in all components of the
Galaxy, including red giants in globular clusters. The data reduction
pipeline will automatically determine 14 elemental abundances,
including overall [Fe/H] metallicity, $\alpha$ elements, and most of
the light elements that vary in globular clusters. It will provide a
database for\linebreak studying cluster light-element variations that is
unparalleled in sample size, amount of abundance information per star,
and start-to-finish homogeneity, and ought to shed significant light
on many aspects of cluster light-element abundance variations.

\section{Evolution of the Galactic globular cluster system}
Although efforts have been made to understand the presence or degree
of light-element abundance variations as a function of present-day
globular cluster properties, correlations with present-day mass,
concentration or ellipticity are loose at best. While we expect the
total mass or central density of a cluster during the formation of the
first and second generations to have an influence on its ability to
self-enrich, those properties have clearly evolved significantly over
each cluster's lifetime.

\subsection{Self-enrichment and escape velocity}
One of the more perplexing elements of the question of globular cluster
light-element self-enrichment is the fact\linebreak that most of the Galactic
globular cluster population is unable to retain AGB or massive-star
winds at the present day. This is observable both in the lack of
intracluster material
in globular clusters (Evans et al. 2003\nocite{ES03}; van Loon
et al. 2006\nocite{vS06}; Boyer et al. 2006\nocite{BW06}), and in low present-day escape
velocities. The census of Galactic globular clusters conducted by
McLaughlin \& van der Marel (2005)\nocite{Mv05} includes values for $v_{esc}$, and roughly half of the
clusters have $v_{esc}$ below $20 {\rm km s}^{-1}$. Since these clusters all have
light-element abundance variations, and since all proposed sources of
feedback material have wind speeds $\ge 20 {\rm km s}^{-1}$, these clusters must
have had higher escape velocities in the past. The massive first
generation in the D'Ercole et al. (2008)\nocite{DVD08} model provides one natural solution to
this problem, as do suggestions (e.g., Palou{\v s} et al. 2009\nocite{PW09}; Sills \& Glebbeek 2010\nocite{SG10}) that collisions between
winds from multiple stars should result in a lower bulk wind velocity,
trapping wind material that otherwise would escape in the dense inner regions of proto-globular
clusters. It seems clear that the Galactic globular cluster population
has evolved strongly since its formation, both in terms of the overall
cluster mass function (e.g., Parmentier \& Gilmore 2005\nocite{PG05}) and in the structural
properties of individual clusters (e.g., de Marchi et al. 2010\nocite{dMP10}).

\subsection{The initial cluster mass function}
It is curious that light-element abundance variations
are apparently universal among present-day Galactic globular\linebreak clusters,
considering that the initial cluster mass function included many
low-mass clusters that should not have been able to self-enrich according to
current globular cluster formation models. There are two possible
explanations for this coincidence that are quite simple: that
self-enrichment in globular clusters is very common, and occurs at
lower cluster masses than we expect, or that cluster dissolution was
extremely effective early in the lifetime of the Milky Way, with only
a small percentage of the highest-mass clusters surviving to the
present day. Globular cluster formation scenarios that rely on
significant gas infall
tend to promote the first explanation, allowing clusters with
lower-mass first generations to form a second stellar generation. The
numerical study of Marks \& Kroups (2010)\nocite{MK10} found that the expulsion of residual
gas following star formation is very effective at destroying globular
clusters with low initial masses and concentrations. This result both
supports the second explanation and implies that globular clusters
have contributed significant numbers of stars with first-generation
abundances to the construction of the stellar halo of the Milky Way,
as is also suggested by the result of Martell \& Grebel (2010)\nocite{MG10}.

If it is simply coincidental that the minimum mass for a globular
cluster to survive to the present day in the Milky Way is larger than
the minimum mass for a globular cluster forming in the Milky Way to
host two stellar generations, then it is instructive to consider
environments where those conditions are not met. In galactic
environments that are more hospitable to long-lived low-mass globular
clusters, the present-day cluster populations ought to include Milky
Way-like, high-mass, two-population clusters along with lower-mass,
chemically homogeneous globular clusters. In galactic environments
in which it is more difficult for clusters to self-enrich, there would
be some fraction of high-mass globular clusters with homogeneous
light-element\linebreak abundances. Regarding the first possibility, the
theoretical study of Conroy \& Spergel (2011)\nocite{CS10} suggests that inter-\linebreak mediate-aged
clusters in the Large Magellanic Cloud with masses between
$10^{4}{\rm M}_{\odot}$ and $10^{5}{\rm M}_{\odot}$ should be able to retain
first-generation winds and self-enrich because of the relatively low
ram pressure they experience. This claim is bolstered by the
observational study of Milone et al. (2009)\nocite{MBP09}, in which clearly broad and/or
bifurcated main-sequence turn-\linebreak offs were found to be common in
intermediate-aged LMC clusters.

\section{Future challenges}

In order to correctly interpret the photometric complexities observed
in some globular clusters (e.g., Marino et al. 2008\nocite{MVP08}; Milone et
al. 2008\nocite{MBP08}; Han et al. 2009\nocite{HL09}; Lardo et al. 2011\nocite{LBP11}), we must understand the photometric shifts caused by changes in light-element abundances and helium, in addition to those caused by age, overall metallicity and [$\alpha$/Fe]. Current theoretical isochrones (e.g., Bertelli et al. 2008\nocite{BG08}; Dotter et al. 2008\nocite{DC08}; Han et al. 2009\nocite{HK09}) are built from stellar models that allow variations in age, overall metallicity, and sometimes the abundances of $\alpha$-elements and helium. Considering the correlations between light-element abundances and $U$-band photometry reported by, e.g., Marino et al. (2008)\nocite{MVP08}, it seems prudent to expand the theoretical grid of stellar models to test for photometric sensitivity to light-element abundance variations. The study of Dotter et al. (2007)\nocite{DCF07} considered exactly this question, constructing isochrones with enhancements in one of C, N, O, Ne, Mg, Si, S, Ca, Ti, or Fe while maintaining a constant overall heavy-element abundance Z in order to explore the effects of individual-element abundance variations on stellar structure. They find that enhancement in C, N or O abundance caused the isochrones to shift to the blue and reduced main-sequence lifetimes by as much as $15\%$, while an enhanced Mg abundance caused iso-\linebreak chrones to be redder but had a minimal positive effect on main-sequence lifetimes. They did not calculate isochrones for the anticorrelated light-element abundance pattern found in globular clusters, but such an exercise would be\linebreak extremely helpful to our understanding of photometric complexity in globular clusters.

It will also be important to understand whether photometric variations are a generic result of light-element abundance variations, or if not, which globular cluster properties permit or prohibit them from being observed. As an example, large variations in CN and CH bandstrength are almost certainly responsible for $U$-band variations among red giants in relatively high-metallicity ([Fe/H $\ge -1.5$) globular clusters, but not all relatively high-metallicity clusters are known to have complex $U-B$,$B$ CMDs. Additionally, multiplicities in different regions of the CMD do not always correspond. For instance, the cluster $\omega$ Cen has three main sequences and five distinct subgiant branches (Villanova et al. 2007)\nocite{VPK07}, making it unclear how many distinct populations it contains. A search by Piotto (2009)\nocite{P09} of archival HST/ACS photometry uncovered multiple turnoffs in several clusters, and further searches for UV-blue photometry of globular cluster stars in public databases (as done in SDSS by Lardo et al. 2011\nocite{LBP11}) or observatory archives could be a quick and profitable way to confirm or deny the presence of photometric complexity in a large number of Galactic globular clusters.

Developing tools for interpreting integrated spectra of extragalactic globular clusters will dramatically expand our ability to study the effects of cosmological environment on globular cluster formation and self-enrichment. Methods for deriving ages and mean elemental abundances from low-resolution spectra have been adapted from galactic stellar populations studies (e.g., Puzia et al. 2006\nocite{PKG06}; Schiavon 2007\nocite{S07}), and techniques for extracting mean abundances from high-resolution integrated spectra of extragalactic globular clusters have been developed by Colucci et al. (2009)\nocite{CB09}. A merger of the two approaches, matching high-resolution spectroscopic data to synthetic spectra that are a sum over multiple distinct populations, will allow detailed searches for abundance variations in extragalactic globular clusters to very large distances.

It is becoming an accepted paradigm that the majority of, if not all, ``normal'' Galactic globular clusters contain stars with a range of light-element abundances, although they are resolutely mono-metallic. This requires that clearly multi-metallic, multi-age clusters like $\omega$ Cen and M54\linebreak formed in different environments, and not as subsystems of the Milky Way. Rather, their extended star formation histories and ability to retain supernova feedback indicate that their early development occurred in a fairly high-mass environment. M54 lies quite close to the core of the Sagittarius dwarf galaxy, prompting some to claim that it formed as the nucleus of the galaxy (e.g., Layden \& Sarajedini 2000\nocite{LS00}), while others argue that M54 formed as a normal globular cluster but is being trapped in the galactic nucleus (e.g., Bellazzini et al. 2008\nocite{BIC08}). One group (Carretta et al. 2010a)\nocite{CBG10} has made the claim that M54 and $\omega$ Cen both originated as nuclear star clusters in dwarf galaxies, with $\omega$ Cen having been captured by the Milky Way earlier while M54 is still being removed from its galaxy of origin. The schematic model of multi-metallicity globular clusters having formed as nuclear star clusters in dwarf galaxies (e.g., Georgiev et al. 2009\nocite{GHP09}) is attractive: the dark-matter halo of the galaxy would permit the cluster to experience extended feedback and star formation, and present-day nuclear star clusters are similar to multi-metallicity globular clusters in several properties such as half-light radius, escape velocity and horizontal branch morphology. 

Recent announcements of mild [Fe/H] and [Ca/Fe] variations in NGC 2419 (Cohen et al. 2010)\nocite{CK10}, along with the discovery of photometric complexity (which may be a result of age or metallicity variations) in several otherwise unexceptional globular clusters (e.g., Piotto 2009\nocite{P09}), raise the question of whether there is a class of globular clusters intermediate between ``normal'' mono-metallic, light element-variable globular clusters and the more massive multi-\linebreak metallicity clusters. Theoretical studies of supernova feedback in extremely massive proto-globular clusters would help to clarify the feasibility of claiming that clusters with mild metallicity variations constitute the high-mass end of typical globular cluster self-enrichment. Numerical simulations of interactions between nucleated dwarf galaxies and the Milky Way would provide an estimate of how many nuclear star clusters may have been captured into Milky Way orbit, and whether clusters like NGC 2419 can be considered as examples of captured nuclear star clusters with a history of low-efficiency feedback. 

\acknowledgements{SLM would like to thank the Scientific Organizing Committee of the Astronomische Gesellschaft 2010 Annual Meeting for the invitation to speak on this subject.}

%\newpage%%%%%%%%%%%%%%%%%%%%%%%%%%%%%%%%%%%%%%%%%%%%%%%%%%%%%%

%\bibliography{ag}

\begin{thebibliography}{106}
\expandafter\ifx\csname natexlab\endcsname\relax\def\natexlab#1{#1}\fi
\expandafter\ifx\csname href\endcsname\relax
  \def\href#1#2{}\fi
\expandafter\ifx\csname urllinklabel\endcsname\relax
  \def\urllinklabel{[LINK]}\fi
\expandafter\ifx\csname adsurllinklabel\endcsname\relax
  \def\adsurllinklabel{[ADS]}\fi

\bibitem[{{Allende Prieto} {et~al.}(2008){Allende Prieto}, {Majewski},
  {Schiavon}, {Cunha}, {Frinchaboy}, {Holtzman}, {Johnston}, {Shetrone},
  {Skrutskie}, {Smith}, \& {Wilson}}]{AP08}
{Allende Prieto}, C., {Majewski}, S.~R., {Schiavon}, R., {Cunha}, K.,
  {Frinchaboy}, P., {Holtzman}, J., {Johnston}, K., {Shetrone}, M.,
  {Skrutskie}, M., {Smith}, V., \& {Wilson}, J. 2008, Astronomische
  Nachrichten, 329, 1018


\bibitem[{{Anderson} {et~al.}(2009){Anderson}, {Piotto}, {King}, {Bedin}, \&
  {Guhathakurta}}]{APK09}
{Anderson}, J., {Piotto}, G., {King}, I.~R., {Bedin}, L.~R., \& {Guhathakurta},
  P. 2009, \apjl, 697, L58


\bibitem[{{Arp} \& {Johnson}(1955)}]{AJ55}
{Arp}, H.~C. \& {Johnson}, H.~L. 1955, \apj, 122, 171


\bibitem[{{Bedin} {et~al.}(2004){Bedin}, {Piotto}, {Anderson}, {Cassisi},
  {King}, {Momany}, \& {Carraro}}]{BPA04}
{Bedin}, L.~R., {Piotto}, G., {Anderson}, J., {Cassisi}, S., {King}, I.~R.,
  {Momany}, Y., \& {Carraro}, G. 2004, \apjl, 605, L125


\bibitem[{{Bell} \& {Dickens}(1980)}]{BD80}
{Bell}, R.~A. \& {Dickens}, R.~J. 1980, \apj, 242, 657


\bibitem[{{Bellazzini} {et~al.}(2008){Bellazzini}, {Ibata}, {Chapman},
  {Mackey}, {Monaco}, {Irwin}, {Martin}, {Lewis}, \& {Dalessandro}}]{BIC08}
{Bellazzini}, M., {Ibata}, R.~A., {Chapman}, S.~C., {Mackey}, A.~D., {Monaco},
  L., {Irwin}, M.~J., {Martin}, N.~F., {Lewis}, G.~F., \& {Dalessandro}, E.
  2008, \aj, 136, 1147


\bibitem[{{Bellini} {et~al.}(2010){Bellini}, {Bedin}, {Piotto}, {Milone},
  {Marino}, \& {Villanova}}]{BBP10}
{Bellini}, A., {Bedin}, L.~R., {Piotto}, G., {Milone}, A.~P., {Marino}, A.~F.,
  \& {Villanova}, S. 2010, \aj, 140, 631


\bibitem[{{Bertelli} {et~al.}(2008){Bertelli}, {Girardi}, {Marigo}, \&
  {Nasi}}]{BG08}
{Bertelli}, G., {Girardi}, L., {Marigo}, P., \& {Nasi}, E. 2008, \aaa, 484, 815


\bibitem[{{Bogdanov} {et~al.}(2010){Bogdanov}, {van den Berg}, {Heinke},
  {Cohn}, {Lugger}, \& {Grindlay}}]{BvH10}
{Bogdanov}, S., {van den Berg}, M., {Heinke}, C.~O., {Cohn}, H.~N., {Lugger},
  P.~M., \& {Grindlay}, J.~E. 2010, \apj, 709, 241


\bibitem[{{Boudreault} \& {Caballero}(2010)}]{BC10}
{Boudreault}, S. \& {Caballero}, J.~A. 2010, ArXiv e-prints


\bibitem[{{Boyer} {et~al.}(2006){Boyer}, {Woodward}, {van Loon}, {Gordon},
  {Evans}, {Gehrz}, {Helton}, \& {Polomski}}]{BW06}
{Boyer}, M.~L., {Woodward}, C.~E., {van Loon}, J.~T., {Gordon}, K.~D., {Evans},
  A., {Gehrz}, R.~D., {Helton}, L.~A., \& {Polomski}, E.~F. 2006, \aj, 132,
  1415


\bibitem[{{Briley} {et~al.}(2004){Briley}, {Cohen}, \& {Stetson}}]{BCS04}
{Briley}, M.~M., {Cohen}, J.~G., \& {Stetson}, P.~B. 2004, \aj, 127, 1579


\bibitem[{{Butler} {et~al.}(1975){Butler}, {Carbon}, \& {Kraft}}]{BCK75}
{Butler}, D., {Carbon}, D., \& {Kraft}, R.~P. 1975, in Bulletin of the American
  Astronomical Society, Vol.~7, Bulletin of the American Astronomical Society,
  239--+


\bibitem[{{Canterna}(1976)}]{C76}
{Canterna}, R. 1976, \aj, 81, 228


\bibitem[{{Carretta} {et~al.}(2009{\natexlab{a}}){Carretta}, {Bragaglia},
  {Gratton}, {D'Orazi}, \& {Lucatello}}]{CBG09b}
{Carretta}, E., {Bragaglia}, A., {Gratton}, R., {D'Orazi}, V., \& {Lucatello},
  S. 2009{\natexlab{a}}, \aaa, 508, 695


\bibitem[{{Carretta} {et~al.}(2010{\natexlab{a}}){Carretta}, {Bragaglia},
  {Gratton}, {Lucatello}, {Bellazzini}, {Catanzaro}, {Leone}, {Momany},
  {Piotto}, \& {D'Orazi}}]{CBG10}
{Carretta}, E., {Bragaglia}, A., {Gratton}, R.~G., {Lucatello}, S.,
  {Bellazzini}, M., {Catanzaro}, G., {Leone}, F., {Momany}, Y., {Piotto}, G.,
  \& {D'Orazi}, V. 2010{\natexlab{a}}, \apjl, 714, L7


\bibitem[{{Carretta} {et~al.}(2009{\natexlab{b}}){Carretta}, {Bragaglia},
  {Gratton}, {Lucatello}, {Catanzaro}, {Leone}, {Bellazzini}, {Claudi},
  {D'Orazi}, {Momany}, {Ortolani}, {Pancino}, {Piotto}, {Recio-Blanco}, \&
  {Sabbi}}]{CBG09}
{Carretta}, E., {Bragaglia}, A., {Gratton}, R.~G., {Lucatello}, S.,
  {Catanzaro}, G., {Leone}, F., {Bellazzini}, M., {Claudi}, R., {D'Orazi}, V.,
  {Momany}, Y., {Ortolani}, S., {Pancino}, E., {Piotto}, G., {Recio-Blanco},
  A., \& {Sabbi}, E. 2009{\natexlab{b}}, \aaa, 505, 117


\bibitem[{{Carretta} {et~al.}(2010{\natexlab{b}}){Carretta}, {Bragaglia},
  {Gratton}, {Recio-Blanco}, {Lucatello}, {D'Orazi}, \& {Cassisi}}]{CBG10b}
{Carretta}, E., {Bragaglia}, A., {Gratton}, R.~G., {Recio-Blanco}, A.,
  {Lucatello}, S., {D'Orazi}, V., \& {Cassisi}, S. 2010{\natexlab{b}}, \aaa,
  516, A55+


\bibitem[{{Cassisi} {et~al.}(2008){Cassisi}, {Salaris}, {Pietrinferni},
  {Piotto}, {Milone}, {Bedin}, \& {Anderson}}]{CSP08}
{Cassisi}, S., {Salaris}, M., {Pietrinferni}, A., {Piotto}, G., {Milone},
  A.~P., {Bedin}, L.~R., \& {Anderson}, J. 2008, \apjl, 672, L115


\bibitem[{{Cohen}(1978)}]{C78}
{Cohen}, J.~G. 1978, \apj, 223, 487


\bibitem[{{Cohen} {et~al.}(2010){Cohen}, {Kirby}, {Simon}, \& {Geha}}]{CK10}
{Cohen}, J.~G., {Kirby}, E.~N., {Simon}, J.~D., \& {Geha}, M. 2010, ArXiv
  e-prints


\bibitem[{{Colucci} {et~al.}(2009){Colucci}, {Bernstein}, {Cameron},
  {McWilliam}, \& {Cohen}}]{CB09}
{Colucci}, J.~E., {Bernstein}, R.~A., {Cameron}, S., {McWilliam}, A., \&
  {Cohen}, J.~G. 2009, \apj, 704, 385


\bibitem[{{Conroy} \& {Spergel}(2011)}]{CS10}
{Conroy}, C. \& {Spergel}, D.~N. 2011, \apj, 726, 36


\bibitem[{{Cottrell} \& {Da Costa}(1981)}]{CD81}
{Cottrell}, P.~L. \& {Da Costa}, G.~S. 1981, \apjl, 245, L79


\bibitem[{{D'Antona} {et~al.}(1983){D'Antona}, {Gratton}, \& {Chieffi}}]{D83}
{D'Antona}, F., {Gratton}, R., \& {Chieffi}, A. 1983, \memsai, 54, 173


\bibitem[{{De Marchi} {et~al.}(2010){De Marchi}, {Paresce}, \& {Portegies
  Zwart}}]{dMP10}
{De Marchi}, G., {Paresce}, F., \& {Portegies Zwart}, S. 2010, \apj, 718, 105


\bibitem[{{de Mink} {et~al.}(2009){de Mink}, {Pols}, {Langer}, \&
  {Izzard}}]{DM09}
{de Mink}, S.~E., {Pols}, O.~R., {Langer}, N., \& {Izzard}, R.~G. 2009, \aaa,
  507, L1


\bibitem[{{Decressin} {et~al.}(2007{\natexlab{a}}){Decressin}, {Charbonnel}, \&
  {Meynet}}]{DC07}
{Decressin}, T., {Charbonnel}, C., \& {Meynet}, G. 2007{\natexlab{a}}, \aaa,
  475, 859


\bibitem[{{Decressin} {et~al.}(2007{\natexlab{b}}){Decressin}, {Meynet},
  {Charbonnel}, {Prantzos}, \& {Ekstr{\"o}m}}]{DM07}
{Decressin}, T., {Meynet}, G., {Charbonnel}, C., {Prantzos}, N., \&
  {Ekstr{\"o}m}, S. 2007{\natexlab{b}}, \aaa, 464, 1029


\bibitem[{{D'Ercole} {et~al.}(2010){D'Ercole}, {D'Antona}, {Ventura},
  {Vesperini}, \& {McMillan}}]{DDV10}
{D'Ercole}, A., {D'Antona}, F., {Ventura}, P., {Vesperini}, E., \& {McMillan},
  S.~L.~W. 2010, \mnras, 407, 854


\bibitem[{{D'Ercole} {et~al.}(2008){D'Ercole}, {Vesperini}, {D'Antona},
  {McMillan}, \& {Recchi}}]{DVD08}
{D'Ercole}, A., {Vesperini}, E., {D'Antona}, F., {McMillan}, S.~L.~W., \&
  {Recchi}, S. 2008, \mnras, 391, 825


\bibitem[{{Dickens} \& {Bell}(1976)}]{DB76}
{Dickens}, R.~J. \& {Bell}, R.~A. 1976, \apj, 207, 506


\bibitem[{{D'Orazi} \& {Marino}(2010)}]{DM10}
{D'Orazi}, V. \& {Marino}, A.~F. 2010, \apjl, 716, L166


\bibitem[{{Dotter} {et~al.}(2007){Dotter}, {Chaboyer}, {Ferguson}, {Lee},
  {Worthey}, {Jevremovi{\'c}}, \& {Baron}}]{DCF07}
{Dotter}, A., {Chaboyer}, B., {Ferguson}, J.~W., {Lee}, H., {Worthey}, G.,
  {Jevremovi{\'c}}, D., \& {Baron}, E. 2007, \apj, 666, 403


\bibitem[{{Dotter} {et~al.}(2008){Dotter}, {Chaboyer}, {Jevremovi{\'c}},
  {Kostov}, {Baron}, \& {Ferguson}}]{DC08}
{Dotter}, A., {Chaboyer}, B., {Jevremovi{\'c}}, D., {Kostov}, V., {Baron}, E.,
  \& {Ferguson}, J.~W. 2008, \apjs, 178, 89


\bibitem[{{Evans} {et~al.}(2003){Evans}, {Stickel}, {van Loon}, {Eyres},
  {Hopwood}, \& {Penny}}]{ES03}
{Evans}, A., {Stickel}, M., {van Loon}, J.~T., {Eyres}, S.~P.~S., {Hopwood},
  M.~E.~L., \& {Penny}, A.~J. 2003, \aaa, 408, L9


\bibitem[{{Gennaro} {et~al.}(2010){Gennaro}, {Brandner}, {Stolte}, \&
  {Henning}}]{GB10}
{Gennaro}, M., {Brandner}, W., {Stolte}, A., \& {Henning}, T. 2010, ArXiv
  e-prints


\bibitem[{{Georgiev} {et~al.}(2009){Georgiev}, {Hilker}, {Puzia}, {Goudfrooij},
  \& {Baumgardt}}]{GHP09}
{Georgiev}, I.~Y., {Hilker}, M., {Puzia}, T.~H., {Goudfrooij}, P., \&
  {Baumgardt}, H. 2009, \mnras, 396, 1075


\bibitem[{{Gratton} {et~al.}(2000){Gratton}, {Sneden}, {Carretta}, \&
  {Bragaglia}}]{GSCB00}
{Gratton}, R.~G., {Sneden}, C., {Carretta}, E., \& {Bragaglia}, A. 2000, \aaa,
  354, 169


\bibitem[{{Han} {et~al.}(2009{\natexlab{a}}){Han}, {Kim}, {Lee}, {Yi}, {Kim},
  \& {Demarque}}]{HK09}
{Han}, S., {Kim}, Y., {Lee}, Y., {Yi}, S.~K., {Kim}, D., \& {Demarque}, P.
  2009{\natexlab{a}}, {New Yonsei-Yale (Y $^{2}$) Isochrones and
  Horizontal-Branch Evolutionary Tracks with Helium Enhancements}, ed.
  {Richtler, T.~\& Larsen, S.}, 33--+


\bibitem[{{Han} {et~al.}(2009{\natexlab{b}}){Han}, {Lee}, {Joo}, {Sohn},
  {Yoon}, {Kim}, \& {Lee}}]{HL09}
{Han}, S., {Lee}, Y., {Joo}, S., {Sohn}, S.~T., {Yoon}, S., {Kim}, H., \&
  {Lee}, J. 2009{\natexlab{b}}, \apjl, 707, L190


\bibitem[{{Harbeck} {et~al.}(2003){Harbeck}, {Smith}, \& {Grebel}}]{HSG03}
{Harbeck}, D., {Smith}, G.~H., \& {Grebel}, E.~K. 2003, \aj, 125, 197


\bibitem[{{Hesser} {et~al.}(1982){Hesser}, {Bell}, {Harris}, \&
  {Cannon}}]{HBH82}
{Hesser}, J.~E., {Bell}, R.~A., {Harris}, G.~L.~H., \& {Cannon}, R.~D. 1982,
  \aj, 87, 1470


\bibitem[{{Iben}(1965)}]{I65}
{Iben}, Jr., I. 1965, \apj, 142, 1447


\bibitem[{{Iben}(1975)}]{I75}
---. 1975, \apj, 196, 525


\bibitem[{{Kayser} {et~al.}(2008){Kayser}, {Hilker}, {Grebel}, \&
  {Willemsen}}]{KHG08}
{Kayser}, A., {Hilker}, M., {Grebel}, E.~K., \& {Willemsen}, P.~G. 2008, \aaa,
  486, 437


\bibitem[{{Kraft} \& {Ivans}(2003)}]{KI03}
{Kraft}, R.~P. \& {Ivans}, I.~I. 2003, \pasp, 115, 143


\bibitem[{{Kroupa} {et~al.}(1993){Kroupa}, {Tout}, \& {Gilmore}}]{KTG93}
{Kroupa}, P., {Tout}, C.~A., \& {Gilmore}, G. 1993, \mnras, 262, 545


\bibitem[{{Langer}(1985)}]{L85}
{Langer}, G.~E. 1985, \pasp, 97, 382


\bibitem[{{Langer} {et~al.}(1992){Langer}, {Suntzeff}, \& {Kraft}}]{LSK92}
{Langer}, G.~E., {Suntzeff}, N.~B., \& {Kraft}, R.~P. 1992, \pasp, 104, 523


\bibitem[{{Lardo} {et~al.}(2011){Lardo}, {Bellazzini}, {Pancino}, {Carretta},
  {Bragaglia}, \& {Dalessandro}}]{LBP11}
{Lardo}, C., {Bellazzini}, M., {Pancino}, E., {Carretta}, E., {Bragaglia}, A.,
  \& {Dalessandro}, E. 2011, \aaa, 525, A114+


\bibitem[{{Layden} \& {Sarajedini}(2000)}]{LS00}
{Layden}, A.~C. \& {Sarajedini}, A. 2000, \aj, 119, 1760


\bibitem[{{Letarte} {et~al.}(2010){Letarte}, {Hill}, {Tolstoy}, {Jablonka},
  {Shetrone}, {Venn}, {Spite}, {Irwin}, {Battaglia}, {Helmi}, {Primas},
  {Francois}, {Kaufer}, {Szeifert}, {Arimoto}, \& {Sadakane}}]{LHT10}
{Letarte}, B., {Hill}, V., {Tolstoy}, E., {Jablonka}, P., {Shetrone}, M.,
  {Venn}, K.~A., {Spite}, M., {Irwin}, M.~J., {Battaglia}, G., {Helmi}, A.,
  {Primas}, F., {Francois}, P., {Kaufer}, A., {Szeifert}, T., {Arimoto}, N., \&
  {Sadakane}, K. 2010, ArXiv e-prints


\bibitem[{{Lorimer}(2010)}]{L10}
{Lorimer}, D.~R. 2010, ArXiv e-prints


\bibitem[{{Mallia}(1975)}]{M75}
{Mallia}, E.~A. 1975, \mnras, 170, 57P


\bibitem[{{Marino} {et~al.}(2008){Marino}, {Villanova}, {Piotto}, {Milone},
  {Momany}, {Bedin}, \& {Medling}}]{MVP08}
{Marino}, A.~F., {Villanova}, S., {Piotto}, G., {Milone}, A.~P., {Momany}, Y.,
  {Bedin}, L.~R., \& {Medling}, A.~M. 2008, \aaa, 490, 625


\bibitem[{{Marks} \& {Kroupa}(2010)}]{MK10}
{Marks}, M. \& {Kroupa}, P. 2010, \mnras, 406, 2000


\bibitem[{{Martell} \& {Grebel}(2010)}]{MG10}
{Martell}, S.~L. \& {Grebel}, E.~K. 2010, \aaa, 519, A14+


\bibitem[{{McClure}(1973)}]{M73}
{McClure}, R.~D. 1973, in IAU Symposium, Vol.~50, Spectral Classification and
  Multicolour Photometry, ed. {C.~Fehrenbach \& B.~E.~Westerlund}, 162--+


\bibitem[{{McClure}(1979)}]{M79}
{McClure}, R.~D. 1979, \memsai, 50, 15


\bibitem[{{McLaughlin} \& {van der Marel}(2005)}]{Mv05}
{McLaughlin}, D.~E. \& {van der Marel}, R.~P. 2005, \apjs, 161, 304


\bibitem[{{McWilliam} \& {Smecker-Hane}(2005)}]{MWSH05}
{McWilliam}, A. \& {Smecker-Hane}, T.~A. 2005, in Astronomical Society of the
  Pacific Conference Series, Vol. 336, Cosmic Abundances as Records of Stellar
  Evolution and Nucleosynthesis, ed. {T.~G.~Barnes III \& F.~N.~Bash}, 221--+


\bibitem[{{Milone} {et~al.}(2009){Milone}, {Bedin}, {Piotto}, \&
  {Anderson}}]{MBP09}
{Milone}, A.~P., {Bedin}, L.~R., {Piotto}, G., \& {Anderson}, J. 2009, \aaa,
  497, 755


\bibitem[{{Milone} {et~al.}(2008){Milone}, {Bedin}, {Piotto}, {Anderson},
  {King}, {Sarajedini}, {Dotter}, {Chaboyer}, {Mar{\'{\i}}n-Franch},
  {Majewski}, {Aparicio}, {Hempel}, {Paust}, {Reid}, {Rosenberg}, \&
  {Siegel}}]{MBP08}
{Milone}, A.~P., {Bedin}, L.~R., {Piotto}, G., {Anderson}, J., {King}, I.~R.,
  {Sarajedini}, A., {Dotter}, A., {Chaboyer}, B., {Mar{\'{\i}}n-Franch}, A.,
  {Majewski}, S., {Aparicio}, A., {Hempel}, M., {Paust}, N.~E.~Q., {Reid},
  I.~N., {Rosenberg}, A., \& {Siegel}, M. 2008, \apj, 673, 241


\bibitem[{{Milone} {et~al.}(2010){Milone}, {Piotto}, {King}, {Bedin},
  {Anderson}, {Marino}, {Momany}, {Malavolta}, \& {Villanova}}]{MPK10}
{Milone}, A.~P., {Piotto}, G., {King}, I.~R., {Bedin}, L.~R., {Anderson}, J.,
  {Marino}, A.~F., {Momany}, Y., {Malavolta}, L., \& {Villanova}, S. 2010,
  \apj, 709, 1183


\bibitem[{{Norris}(1978)}]{N78}
{Norris}, J. 1978, in IAU Symposium, Vol.~80, The HR Diagram - The 100th
  Anniversary of Henry Norris Russell, ed. {A.~G.~D.~Philip \& D.~S.~Hayes},
  195--203


\bibitem[{{Norris}(1987)}]{N87}
{Norris}, J. 1987, \apjl, 313, L65


\bibitem[{{Norris} \& {Cottrell}(1979)}]{NC79}
{Norris}, J. \& {Cottrell}, P.~L. 1979, \apjl, 229, L69


\bibitem[{{Norris} {et~al.}(1981){Norris}, {Cottrell}, {Freeman}, \& {Da
  Costa}}]{N81}
{Norris}, J., {Cottrell}, P.~L., {Freeman}, K.~C., \& {Da Costa}, G.~S. 1981,
  \apj, 244, 205


\bibitem[{{Norris} {et~al.}(1984){Norris}, {Freeman}, \& {Da Costa}}]{N84}
{Norris}, J., {Freeman}, K.~C., \& {Da Costa}, G.~S. 1984, \apj, 277, 615


\bibitem[{{Norris} \& {Zinn}(1977)}]{NZ77}
{Norris}, J. \& {Zinn}, R. 1977, \apj, 215, 74


\bibitem[{{Palou{\v s}} {et~al.}(2009){Palou{\v s}}, {W{\"u}nsch},
  {Tenorio-Tagle}, \& {Silich}}]{PW09}
{Palou{\v s}}, J., {W{\"u}nsch}, R., {Tenorio-Tagle}, G., \& {Silich}, S. 2009,
  in IAU Symposium, Vol. 254, IAU Symposium, ed. {J.~Andersen,
  J.~Bland-Hawthorn, \& B.~Nordstr{\"o}m}, 233--238


\bibitem[{{Pancino} {et~al.}(2010){Pancino}, {Rejkuba}, {Zoccali}, \&
  {Carrera}}]{PR10}
{Pancino}, E., {Rejkuba}, M., {Zoccali}, M., \& {Carrera}, R. 2010, ArXiv
  e-prints


\bibitem[{{Parmentier} \& {Gilmore}(2005)}]{PG05}
{Parmentier}, G. \& {Gilmore}, G. 2005, \mnras, 363, 326


\bibitem[{{Parmentier} {et~al.}(1999){Parmentier}, {Jehin}, {Magain},
  {Neuforge}, {Noels}, \& {Thoul}}]{PJ99}
{Parmentier}, G., {Jehin}, E., {Magain}, P., {Neuforge}, C., {Noels}, A., \&
  {Thoul}, A.~A. 1999, \aaa, 352, 138


\bibitem[{{Peterson}(1980)}]{P80}
{Peterson}, R.~C. 1980, \apjl, 237, L87


\bibitem[{{Pilachowski} {et~al.}(1996{\natexlab{a}}){Pilachowski}, {Sneden}, \&
  {Kraft}}]{PSK96}
{Pilachowski}, C.~A., {Sneden}, C., \& {Kraft}, R.~P. 1996{\natexlab{a}}, \aj,
  111, 1689


\bibitem[{{Pilachowski} {et~al.}(1996{\natexlab{b}}){Pilachowski}, {Sneden},
  {Kraft}, \& {Langer}}]{P96}
{Pilachowski}, C.~A., {Sneden}, C., {Kraft}, R.~P., \& {Langer}, G.~E.
  1996{\natexlab{b}}, \aj, 112, 545


\bibitem[{{Piotto}(2009)}]{P09}
{Piotto}, G. 2009, in IAU Symposium, Vol. 258, IAU Symposium, ed.
  {E.~E.~Mamajek, D.~R.~Soderblom, \& R.~F.~G.~Wyse}, 233--244


\bibitem[{{Piotto} {et~al.}(2007){Piotto}, {Bedin}, {Anderson}, {King},
  {Cassisi}, {Milone}, {Villanova}, {Pietrinferni}, \& {Renzini}}]{PBA07}
{Piotto}, G., {Bedin}, L.~R., {Anderson}, J., {King}, I.~R., {Cassisi}, S.,
  {Milone}, A.~P., {Villanova}, S., {Pietrinferni}, A., \& {Renzini}, A. 2007,
  \apjl, 661, L53


\bibitem[{{Preston}(1961)}]{P61}
{Preston}, G.~W. 1961, \apj, 134, 651


\bibitem[{{Puzia} {et~al.}(2006){Puzia}, {Kissler-Patig}, \&
  {Goudfrooij}}]{PKG06}
{Puzia}, T.~H., {Kissler-Patig}, M., \& {Goudfrooij}, P. 2006, \apj, 648, 383


\bibitem[{{Sandage}(1958)}]{S58}
{Sandage}, A. 1958, Ricerche Astronomiche, 5, 41


\bibitem[{{Sandage} \& {Wallerstein}(1960)}]{SW60}
{Sandage}, A. \& {Wallerstein}, G. 1960, \apj, 131, 598


\bibitem[{{Sbordone} {et~al.}(2007){Sbordone}, {Bonifacio}, {Buonanno},
  {Marconi}, {Monaco}, \& {Zaggia}}]{SBB07}
{Sbordone}, L., {Bonifacio}, P., {Buonanno}, R., {Marconi}, G., {Monaco}, L.,
  \& {Zaggia}, S. 2007, \aaa, 465, 815


\bibitem[{{Schiavon}(2007)}]{S07}
{Schiavon}, R.~P. 2007, \apjs, 171, 146


\bibitem[{{Shen} {et~al.}(2010){Shen}, {Bonifacio}, {Pasquini}, \&
  {Zaggia}}]{SB10}
{Shen}, Z., {Bonifacio}, P., {Pasquini}, L., \& {Zaggia}, S. 2010, \aaa, 524,
  L2+


\bibitem[{{Shetrone} {et~al.}(2003){Shetrone}, {Venn}, {Tolstoy}, {Primas},
  {Hill}, \& {Kaufer}}]{SVT03}
{Shetrone}, M., {Venn}, K.~A., {Tolstoy}, E., {Primas}, F., {Hill}, V., \&
  {Kaufer}, A. 2003, \aj, 125, 684


\bibitem[{{Shetrone}(1996)}]{S96}
{Shetrone}, M.~D. 1996, \aj, 112, 1517


\bibitem[{{Siegel} {et~al.}(2007){Siegel}, {Dotter}, {Majewski}, {Sarajedini},
  {Chaboyer}, {Nidever}, {Anderson}, {Mar{\'{\i}}n-Franch}, {Rosenberg},
  {Bedin}, {Aparicio}, {King}, {Piotto}, \& {Reid}}]{SDM07}
{Siegel}, M.~H., {Dotter}, A., {Majewski}, S.~R., {Sarajedini}, A., {Chaboyer},
  B., {Nidever}, D.~L., {Anderson}, J., {Mar{\'{\i}}n-Franch}, A., {Rosenberg},
  A., {Bedin}, L.~R., {Aparicio}, A., {King}, I., {Piotto}, G., \& {Reid},
  I.~N. 2007, \apjl, 667, L57


\bibitem[{{Sills} \& {Glebbeek}(2010)}]{SG10}
{Sills}, A. \& {Glebbeek}, E. 2010, \mnras, 407, 277


\bibitem[{{Smith}(2008)}]{S08}
{Smith}, G.~H. 2008, \pasp, 120, 952


\bibitem[{{Smith}(2010)}]{S10}
---. 2010, \pasp, 122, 1171


\bibitem[{{Sneden} {et~al.}(1992){Sneden}, {Kraft}, {Prosser}, \&
  {Langer}}]{S92}
{Sneden}, C., {Kraft}, R.~P., {Prosser}, C.~F., \& {Langer}, G.~E. 1992, \aj,
  104, 2121


\bibitem[{{Sollima} {et~al.}(2007){Sollima}, {Ferraro}, {Bellazzini},
  {Origlia}, {Straniero}, \& {Pancino}}]{SFB07}
{Sollima}, A., {Ferraro}, F.~R., {Bellazzini}, M., {Origlia}, L., {Straniero},
  O., \& {Pancino}, E. 2007, \apj, 654, 915


\bibitem[{{Str{\"o}mgren}(1963)}]{S63}
{Str{\"o}mgren}, B. 1963, Quarterly Journal of the Royal Astronomical Society,
  4, 8


\bibitem[{{Suntzeff}(1981)}]{S81}
{Suntzeff}, N.~B. 1981, \apjs, 47, 1


\bibitem[{{Sweigart} \& {Mengel}(1979)}]{SM79}
{Sweigart}, A.~V. \& {Mengel}, J.~G. 1979, \apj, 229, 624


\bibitem[{{van Loon} {et~al.}(2006){van Loon}, {Stanimirovi{\'c}}, {Evans}, \&
  {Muller}}]{vS06}
{van Loon}, J.~T., {Stanimirovi{\'c}}, S., {Evans}, A., \& {Muller}, E. 2006,
  \mnras, 365, 1277


\bibitem[{{Villanova} {et~al.}(2007){Villanova}, {Piotto}, {King}, {Anderson},
  {Bedin}, {Gratton}, {Cassisi}, {Momany}, {Bellini}, {Cool}, {Recio-Blanco},
  \& {Renzini}}]{VPK07}
{Villanova}, S., {Piotto}, G., {King}, I.~R., {Anderson}, J., {Bedin}, L.~R.,
  {Gratton}, R.~G., {Cassisi}, S., {Momany}, Y., {Bellini}, A., {Cool}, A.~M.,
  {Recio-Blanco}, A., \& {Renzini}, A. 2007, \apj, 663, 296


\bibitem[{{Yanny} {et~al.}(2009){Yanny}, {Rockosi}, {Newberg}, {Knapp},
  {Adelman-McCarthy}, {Alcorn}, {Allam}, {Allende Prieto}, {An}, {Anderson},
  {Anderson}, {Bailer-Jones}, {Bastian}, {Beers}, {Bell}, {Belokurov},
  {Bizyaev}, {Blythe}, {Bochanski}, {Boroski}, {Brinchmann}, {Brinkmann},
  {Brewington}, {Carey}, {Cudworth}, {Evans}, {Evans}, {Gates}, {G{\"a}nsicke},
  {Gillespie}, {Gilmore}, {Gomez-Moran}, {Grebel}, {Greenwell}, {Gunn},
  {Jordan}, {Jordan}, {Harding}, {Harris}, {Hendry}, {Holder}, {Ivans},
  {Ivezi{\v c}}, {Jester}, {Johnson}, {Kent}, {Kleinman}, {Kniazev},
  {Krzesinski}, {Kron}, {Kuropatkin}, {Lebedeva}, {Lee}, {Leger}, {L{\'e}pine},
  {Levine}, {Lin}, {Long}, {Loomis}, {Lupton}, {Malanushenko}, {Malanushenko},
  {Margon}, {Martinez-Delgado}, {McGehee}, {Monet}, {Morrison}, {Munn},
  {Neilsen}, {Nitta}, {Norris}, {Oravetz}, {Owen}, {Padmanabhan}, {Pan},
  {Peterson}, {Pier}, {Platson}, {Fiorentin}, {Richards}, {Rix}, {Schlegel},
  {Schneider}, {Schreiber}, {Schwope}, {Sibley}, {Simmons}, {Snedden}, {Smith},
  {Stark}, {Stauffer}, {Steinmetz}, {Stoughton}, {Subba Rao}, {Szalay},
  {Szkody}, {Thakar}, {Thirupathi}, {Tucker}, {Uomoto}, {Vanden Berk},
  {Vidrih}, {Wadadekar}, {Watters}, {Wilhelm}, {Wyse}, {Yarger}, \&
  {Zucker}}]{Y09}
{Yanny}, B., {Rockosi}, C., {Newberg}, H.~J., {Knapp}, G.~R.,
  {Adelman-McCarthy}, J.~K., {Alcorn}, B., {Allam}, S., {Allende Prieto}, C.,
  {An}, D., {Anderson}, K.~S.~J., {Anderson}, S., {Bailer-Jones}, C.~A.~L.,
  {Bastian}, S., {Beers}, T.~C., {Bell}, E., {Belokurov}, V., {Bizyaev}, D.,
  {Blythe}, N., {Bochanski}, J.~J., {Boroski}, W.~N., {Brinchmann}, J.,
  {Brinkmann}, J., {Brewington}, H., {Carey}, L., {Cudworth}, K.~M., {Evans},
  M., {Evans}, N.~W., {Gates}, E., {G{\"a}nsicke}, B.~T., {Gillespie}, B.,
  {Gilmore}, G., {Gomez-Moran}, A.~N., {Grebel}, E.~K., {Greenwell}, J.,
  {Gunn}, J.~E., {Jordan}, C., {Jordan}, W., {Harding}, P., {Harris}, H.,
  {Hendry}, J.~S., {Holder}, D., {Ivans}, I.~I., {Ivezi{\v c}}, {\v Z}.,
  {Jester}, S., {Johnson}, J.~A., {Kent}, S.~M., {Kleinman}, S., {Kniazev}, A.,
  {Krzesinski}, J., {Kron}, R., {Kuropatkin}, N., {Lebedeva}, S., {Lee}, Y.~S.,
  {Leger}, R.~F., {L{\'e}pine}, S., {Levine}, S., {Lin}, H., {Long}, D.~C.,
  {Loomis}, C., {Lupton}, R., {Malanushenko}, O., {Malanushenko}, V., {Margon},
  B., {Martinez-Delgado}, D., {McGehee}, P., {Monet}, D., {Morrison}, H.~L.,
  {Munn}, J.~A., {Neilsen}, E.~H., {Nitta}, A., {Norris}, J.~E., {Oravetz}, D.,
  {Owen}, R., {Padmanabhan}, N., {Pan}, K., {Peterson}, R.~S., {Pier}, J.~R.,
  {Platson}, J., {Fiorentin}, P.~R., {Richards}, G.~T., {Rix}, H., {Schlegel},
  D.~J., {Schneider}, D.~P., {Schreiber}, M.~R., {Schwope}, A., {Sibley}, V.,
  {Simmons}, A., {Snedden}, S.~A., {Smith}, J.~A., {Stark}, L., {Stauffer}, F.,
  {Steinmetz}, M., {Stoughton}, C., {Subba Rao}, M., {Szalay}, A., {Szkody},
  P., {Thakar}, A.~R., {Thirupathi}, S., {Tucker}, D., {Uomoto}, A., {Vanden
  Berk}, D., {Vidrih}, S., {Wadadekar}, Y., {Watters}, S., {Wilhelm}, R.,
  {Wyse}, R.~F.~G., {Yarger}, J., \& {Zucker}, D. 2009, \aj, 137, 4377


\bibitem[{{Yong} \& {Grundahl}(2008)}]{YG08}
{Yong}, D. \& {Grundahl}, F. 2008, \apjl, 672, L29


\bibitem[{{Yong} {et~al.}(2009){Yong}, {Grundahl}, {D'Antona}, {Karakas},
  {Lattanzio}, \& {Norris}}]{YGD09}
{Yong}, D., {Grundahl}, F., {D'Antona}, F., {Karakas}, A.~I., {Lattanzio},
  J.~C., \& {Norris}, J.~E. 2009, \apjl, 695, L62


\bibitem[{{Yong} {et~al.}(2008){Yong}, {Karakas}, {Lambert}, {Chieffi}, \&
  {Limongi}}]{YK08}
{Yong}, D., {Karakas}, A.~I., {Lambert}, D.~L., {Chieffi}, A., \& {Limongi}, M.
  2008, \apj, 689, 1031


\bibitem[{{Zinn}(1973)}]{Z73}
{Zinn}, R. 1973, \aaa, 25, 409


\bibitem[{{Zinn}(1977)}]{Z77}
---. 1977, \apj, 218, 96


\end{thebibliography}

\end{document}